# Cycle-guided Denoising Diffusion Probability Model for 3D Cross-modality MRI Synthesis

Shaoyan Pan[1,2], Chih-Wei Chang[1], Junbo Peng[1], Jiahan Zhang[1], Richard L.J. Qiu[1],

Tonghe Wang[3], Justin Roper[1], Tian Liu[4], Hui Mao[5] and Xiaofeng Yang[1,2*]

[1]Department of Radiation Oncology and Winship Cancer Institute, Emory University, Atlanta, GA 30322, USA

[2]Department of Biomedical Informatics, Emory University, Atlanta, GA 30322, USA

[3]Medical Physics, Memorial Sloan Kettering Cancer Center, New York, NY 10065, USA

[4]Radiation Oncology, Icahn School of Medicine at Mount Sinai, New York, NY 10029, USA

[5]Department of Radiology and Imaging Sciences and Winship Cancer Institute, Atlanta, GA 30308

*Corresponding to: xiaofeng.yang@emory.edu

**Abstract** This study aims to develop a novel Cycle-guided Denoising Diffusion Probability Model (CG-DDPM) for cross-modality MRI synthesis. The CG-DDPM deploys two DDPMs that condition each other to generate synthetic images from two different MRI pulse sequences. The two DDPMs exchange random latent noise in the reverse processes, which helps to regularize both DDPMs and generate matching images in two modalities. This improves image-to-image translation accuracy. We evaluated the CG-DDPM quantitatively using mean absolute error (MAE), multi-scale structural similarity index measure (MSSIM), and peak signal-to-noise ratio (PSNR), as well as the network synthesis consistency, on the BraTS2020 dataset. Our proposed method showed high accuracy and reliable consistency for MRI synthesis. In addition, we compared the CG-DDPM with several other state-of-the-art networks and demonstrated statistically significant improvements in the image quality of synthetic MRIs. The proposed method enhances the capability of current multimodal MRI synthesis approaches, which could contribute to more accurate diagnosis and better treatment planning for patients by synthesizing additional MRI modalities.

## 1    Introduction

Magnetic Resonance Imaging (MRI) is widely deployed in the clinic for diagnosis, prognosis, and treatment planning for radiotherapy [1-5]. MRI is featured in providing anatomical and functional information, allowing physicians to accurately identify lesion regions, malignancy types, and metastasis status. For instance, T1-weighted (T1) MRI scans offer clear contrast resolution between white and gray matter to reveal the tissue boundary in diagnosing brain lesions. T2-weighted (T2) MRI scans are used to identify craniospinal fluid (CSF) in the brain, and the fluid-attenuated inversion recovery (FLAIR) technique can suppress fluid signals to enable lesion detection in proximity to CSF. Combining the strengths of each modality enables a more comprehensive examination of the underlying anatomy and physiology, facilitating disease diagnosis and treatment. However, MRI scanning is time-consuming due to fundamental imaging physics. The time dilation between different MRI scanning sequences can cause motion artifacts, compromising the image quality. For patients allergic to MRI contrast, the diagnosis accuracy may be compromised without using contrast.

Deep learning as a universal approximator [6-12] has been explored for MRI synthesis via various model hierarchies [13, 14]. Generative adversarial networks (GAN) [15-19] have been demonstrated to be effective in generating high-quality medical images across different modalities. Recently, denoising diffusion probability models (DDPM) [20, 21] have gained attraction for image synthesis since DDPM addresses the issues of GAN, such as mode collapse when learning multimodal distribution. However, DDPM requires Markov chain denoising processes, which root in uncertainty for image synthesis [22]. Numerous efforts and studies have been conducted to enhance the performance of DDPM [23, 24] and applied the DDPM into medical image synthesis [25]. In radiotherapy, the investigation of three-dimensional (3D) DDPM's uncertainty on MRI synthesis is still in its infancy, and such a development is essential to ensure patient safety and benefits.

This work proposes a cycle-guided denoising diffusion probability model (CG-DDPM), a cross-modality MRI synthesis algorithm that generates comprehensive MRI modalities based on available inputs. To the authors' best knowledge, the proposed model is the first capable of leveraging the 3D MRI data to ensure the consistency of patient anatomy. In conventional DDPMs, the reverse denoising process is entirely random, leading to inconsistent synthetic results across multiple runs and potentially causing the model to generate MRIs in the target signal sequences that can mismatch those in the source signal sequences. In contrast, we propose a cycle-guided reverse latent process to regularize DDPM to conserve the patient anatomy and improve image translation accuracy. To demonstrate the robustness and accuracy of the CG-DDPM, we investigate its performance with other state-of-the-art MRI cross-modality translation methods in BraTS2020 dataset [26-28]. The originality of the proposed CG-DDPM, the first 3D DDPM-based model for MRI synthesis, can be summarized in two aspects. (The code will be released after acceptance from Journal)

• The proposed 3D CG-DDPM can stably preserve patient anatomy and reproduce reverse noise pattern when synthesizing MRI images. This feature is essential for patient safety regarding lesion diagnosis and prognosis.

• The proposed 3D CG-DDPM can improve accuracy compared to other state-of-the-art GAN and DDPM methods. This outcome is essential for potential MRI-based treatment planning.

## 2 Method: Cycle-Guided Denoising Diffusion Probability Model

The proposed CG-DDPM consists of two identical denoising neural networks run in three processes (Figure 1). Firstly, a forward diffusion process adds a small amount of Gaussian noise to the high-quality, scanner-acquired target MRI $X_0$ and the source MRI $Y_0$ at $N$ successive time step, resulting in two sequences of noisy images $[X_1, ..., X_N]$ and $[Y_1, ..., Y_N]$. Next, a reverse diffusion process trains two neural networks $f^X$, under the guidance of the source MRI $Y_0$, and $f^Y$, under the guidance of the target MRI $X_0$, to denoise the MRI $X_n$ to $X_{n-1}$, and $Y_n$ to $Y_{n-1}$ at any timestep $n$, respectively. In addition, the reverse latent noise $\epsilon_N^X$ obtained between the $X_n$ and $X_{n-1}$ is used in the generation of $Y_{n-1}$, and vice versa, to control the denoising direction. Thirdly, in an inference process, the fine-trained source reverse process $p_f^Y$ generates a sequence of reverse latent code $[z_N, z_{N-1}, ..., z_1]$ from $[Y_0, ..., Y_N]$. The latent codes are then applied to $f^X$ to convert a Gaussian noise $X_N$ to the target MRI $X_{new}$ paired to $Y_0$. In the following sections, we use target image synthesis to illustrate the forward and reverse processes.

### 2.1 Forward diffusion process

The forward diffusion defines the noisy image generation as a Gaussian Markov process $\mathcal{N}\left(X_n; \sqrt{1-\beta_n}X_{n-1}, \beta_n I\right)$, with pre-determined mean of $\sqrt{1-\beta_n}$ and standard deviation of $\beta_n$ at timestep $n$. Following [21], we can efficiently generate a noisy image at any arbitrary timestep $n$ by the clean image $X_0$ of the source or target modality:

$$X_n = \sqrt{\prod_{i=1}^n \alpha_i}\, X_0 + \sqrt{1 - \prod_{i=1}^n \alpha_i}\, \epsilon^X \tag{1}$$

where $\alpha_i \coloneqq 1 - \beta_i$, $\epsilon^X \sim \mathcal{N}(0, I)$ is a noise sampled from a normal distribution.

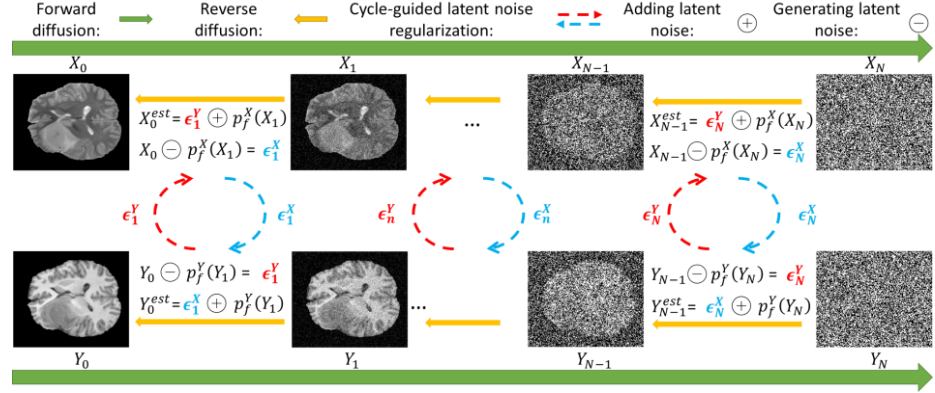

**Figure 1:** Proposed CG-DDPM: In addition to the forward and reverse diffusion in the traditional DDPM, the generation direction is controlled using a cycle-guided latent noise regularization technique. The ground truth target and source MRIs $X_{1,...,N}$ and $Y_{1,...,N}$ are used to generate the deterministic latent noise $\epsilon_{1,...,N}^X$ and $\epsilon_{1,...,N}^Y$. Then we mutually employ the latent noise to generate synthetic MRIs $X_{1,...,N}^{est}$ and $Y_{1,...,N}^{est}$ from two modalities.

### 2. 2 Reverse diffusion process

The reverse diffusion process is defined as another Gaussian process $p^X(X_{n-1}|X_n) \sim \mathcal{N}(X_n; \mu^X, \Sigma^X)$, where $\mu^X$ and $\Sigma^X$ are unknown. We use a neural network $f^X$, conditioned on the clean source image $Y_0$, to estimate them:

$$p_f^X(X_{n-1}|X_n, Y_0, n) = \mathcal{N}\left(X_{n-1}; \mu_f^X(X_n, Y_0, n), \Sigma_f^X(X_n, Y_0, n)\right) \tag{2}$$

Following Nicol et al.'s formulation [23], a denoising network is trained to estimate noise $\epsilon_f^X$ and the variance coefficient $v^X$. The network is optimized as:

$$\underset{f^X}{\text{argmin}}\ L_{DDPM}^X = MAE\left(\epsilon^X, \epsilon_f^X(X_n, Y_0, n)\right) + \gamma * L_{\text{VLB}}(\Sigma^X, \Sigma_f^X(X_n, Y_0, n), n) \quad (3)$$

where $MAE$ is the mean absolute error, $L_{VLB}$ is the variational lower bound loss, and:

$$\Sigma_f^X(X_n, Y_0, n) = \exp\left(v^X(X_n, Y_0, n) * \log\beta_n + (1 - v^X(X_n, Y_0, n)) *\right.$$
$$\left.\log\left(\frac{1 - \prod_{i=1}^{n-1}\alpha_i}{1 - \prod_{i=1}^{n}\alpha_i}\beta_n\right)\right) \quad (4)$$

Then the estimated mean $\mu_f^X$ can be obtained from:

$$\mu_f^X(X_n, Y_0, n) = \frac{1}{\sqrt{\alpha_n}}\left(X_n - \frac{\beta_n}{\sqrt{1 - \alpha_n}}\epsilon_f^X(X_n, Y_0, n)\right) \quad (5)$$

where $\gamma$ is empirically set to 0.05. Full optimization details are in the Appendix. B.

## 2.3 Applying the cycle-guided latent code regularization

With an accurately estimated mean and variance, using the reparameterization trick, we can sample estimated $X_{n-1}^{est}$ from $X_n$, and $Y_{n-1}^{est}$ from $Y_n$ at any timestep $n$:

$$X_{n-1}^{est} = \mu_f^X(X_n, Y_0, n) + \Sigma_f^X(X_n, Y_0, n) * \epsilon_n^X \quad (7)$$

$$Y_{n-1}^{est} = \mu_f^Y(Y_n, X_0, n) + \Sigma_f^Y(Y_n, X_0, n) * \epsilon_n^Y \quad (8)$$

where $\epsilon_n^{X,Y} \sim \mathcal{N}(0, I)$. Previous research [29] suggests that a fixed sequence of random latent codes can produce similar images consistently using two diffusion models. Based on this, our proposed approach is to use the same fixed latent code for both the source and target DDPMs to generate paired MRI. We introduce an additional step in model training, where the network is optimized to generate matching MRI by using the latent code from the reverse process of the source MRI. Formally, the source latent code is:

$$z_n = \epsilon_n^Y = \frac{Y_{n-1} - \mu_f^Y(Y_n, X_0, n)}{\Sigma_f^Y(Y_n, X_0, n)} \quad (9)$$

Then the matching MRI on the target domain is generated as:

$$X_{n-1}^{est} = \mu_f^X(X_n, Y_0, n) + \Sigma_f^X(X_n, Y_0, n) * z_n \quad (20)$$

By applying the same process to source domain reverse process, we jointly optimize the source and target network by the cycle-guided loss $L_{cyc}$, which consists of two $MAE$ losses between the synthetic images and ground truth images:

$$\underset{f^X, f^Y}{\text{argmin}}\ L_{\text{cyc}} = MAE(X_{n-1}^{est}, X_{n-1}) + MAE(Y_{n-1}^{est}, Y_{n-1}) \quad (31)$$

Finally, the CG-DDPM is optimized with a cycle-guided strength $\lambda$ (empirically=1):

$$L_{CG-DDPM} = L_{DDPM}^X + L_{DDPM}^Y + \lambda L_{\text{cyc}} \tag{42}$$

## 2.4 Sampling

To generate a target image matching the given condition image, we firstly generate a sequence of noisy image $[Y_1, \ldots, Y_N]$ based on Eq. (2). Taking the $Y_N$ as a starting point for the target image sampling, we denote $X_N = Y_N$. Then we can obtain the corresponding clean target image $X_0^{est}$:

$$X_0^{est} = \frac{X_n - \sqrt{1 - \prod_{i=1}^N \alpha_i} \epsilon_f^X(X_N, Y_0, N)}{\sqrt{\prod_{i=1}^N \alpha_n} x_0} \tag{53}$$

Then the $X_0^{est}$ can be used for obtaining the first source reverse code at timestep $t$:

$$z_n = \epsilon_N^Y = \frac{Y_{N-1} - \mu_f^Y(Y_N, X_0^{est}, N)}{\Sigma_f^Y(Y_N, X_0^{est}, N)} \tag{64}$$

And we can obtain the less noisy target image $X_{N-1}^{est}$:

$$X_{N-1}^{est} = \mu_f^X(X_N, Y_0, N) + \Sigma_f^X(X_N, Y_0, N) * z_n \tag{75}$$

By repeating the above operation N times, the target DDPM can iteratively generate a less noisy image and eventually obtain a target MRI $X_{new}$ matching the source MRI $Y_0$.

## 2.5 Denoising network

The denoising networks in the CG-DDPM are 3D U-shaped neural networks consisting of convolution layers defined in [30] and Swin-transformer layers defined in [31]. The full architecture is presented in Appendix A.

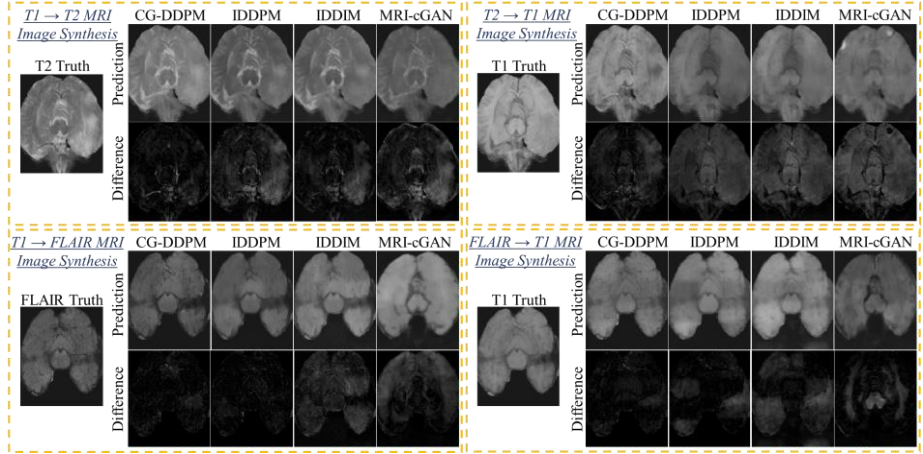

**Figure 2.** Ground truth MRIs and synthetic MRIs for T1→T2, T2→T1, T1→FLAIR, FLAIR→T1 synthesis. Synthetic MRIs from the proposed CG-DDPM (column #1), IDDPM (column #2), IDDIM (column #3), and MRI-cGAN (column #4) are presented column-wise. The difference maps between the truth and synthetic MRIs are shown below.

## 3 Experiments

### 3.1 BraTS 2020 dataset

The BraTS 2020 dataset [26-28] is a collection of multi-institutional multimodal brain MRI scans. For this experiment, we utilized the first 178 scans for training, 2 for validation, and 18 for testing. We resampled each MRI scan into a voxel spacing of 1x1x6 mm, then centered and padded the boundaries, resulting in image grid dimensions of 256x256x32. Each MRI scan is normalized to the intensity interval [-1,1] in both the training and inference phases. During each training iteration, we randomly selected two patches with a size of 64x64x16 from each MRI scan to train the denoising model. During inference, we used a sliding window approach to generate the full MRI scan with a window size equal to the patch size. We set the overlap between windows to 50% of the patch size and applied Gaussian weighting to the edges of the windows. No augmentation or registration techniques were used.

### 3.2 Experiment design

The proposed CG-DDPM was applied for four MRI cross-modality translation tasks using the BraTS 2020 dataset: one model for T1-T2 (T1-to-T2 and T2-to-T1), and one for T1-FLAIR (T1-to- FLAIR and FLAIR-to-T1). Each CG-DDPM was trained with N=4000 forward diffusion steps, and 50 reverse diffusion steps were used to sample synthetic images. The CG-DDPM denoising networks were trained with an AdamW [32] optimizer (initial learning rate of 0.00004 and weight decay rate of 0.001) with a batch size of four. The T1-to-T2 CG-DDPM was trained for 1000 epochs, and the T1-to-FLAIR CG-DDPM was trained for 850 epochs. The source and target denoising networks were trained individually, under Eq. (5), for the first 500 epochs and jointly trained using cycle-guided latent code regularization for the remaining epochs, as described in Eq. (12).

**Table 1.** Quantitative and statistical analysis of the synthetic MRIs for CG-DDPM vs. IDDPM, IDDIM, and MRI-cGAN. MAE is calculated by the normalized images ([-1,1]).

|  |  | CG-DDPM | IDDPM | IDDIM | MRI-cGAN |
|---|---|---|---|---|---|
| T1 | MSSIM | **0.968** | 0.914[*] | 0.882[*] | 0.875[*] |
| ↓ | MAE | **0.011** | 0.013[*] | 0.020[*] | 0.025[*] |
| T2 | PSNR(dB) | 28.6 | **28.8** | 27.4 | 25.2[*] |
| T2 | MSSIM | **0.971** | 0.945[*] | 0.878[*] | 0.924[*] |
| ↓ | MAE | **0.012** | 0.024[*] | 0.024[*] | 0.018[*] |
| T1 | PSNR(dB) | **28.7** | 24.1[*] | 25.1[*] | 26.5[*] |
| T1 | MSSIM | **0.966** | 0.958[*] | 0.881[*] | 0.882[*] |
| ↓ | MAE | **0.011** | 0.012 | 0.022[*] | 0.028[*] |
| FLAIR | PSNR(dB) | **28.8** | 28.2 | 25.6[*] | 23.0[*] |
| FLAIR | MSSIM | **0.971** | 0.957[*] | 0.875[*] | 0.875[*] |
| ↓ | MAE | **0.013** | 0.014 | 0.025[*] | 0.025[*] |
| T1 | PSNR(dB) | 27.7 | **28.1** | 25.2 | 25.2[*] |

[*]Statistically significant difference to CG-DDPM ($p$-value $< 0.05$)

We evaluated the performance of CG-DDPM using three evaluation metrics: multi-scale structural similarity measure (MSSIM) [33], peak signal-to-noise ratio (PSNR), and mean absolute error (MAE). The proposed model was also compared to other deep learning models, including improved DDPM (IDDPM) [23], improved denoising diffusion implicit model (IDDIM) [23], and MRI conditional GAN (MRI-cGAN) [34]. IDDPM and IDDIM were trained for 4000 steps and sampled MRIs for 256 steps based on the settings defined in [23]. Due to the randomness in diffusion-based deep learning models, we used Monte Carlo-based (MC-based) sampling [22] to generate a converged synthetic MRI. We sampled MRIs in five runs and took the average result as the final synthetic MRI for each patient.

### 3.3 Quantitative and qualitative results for accuracy of CG-DDPM

Figure 2 and 3 shows the visualization of synthetic MRIs from various methods. The CG-DDPM and DDPM-based methods exhibit good visual appearance, but the DDPM-based methods display slight patchy artifacts due to the patch-based inference strategy. The computational requirements for full image inference were overwhelming in their hardware setting (48 GB NVIDIA RTX 6000 GPU), therefore patch-based inference is needed. Higher overlapping ratios of the patch-based inference could mitigate the patchy artifacts. Table 1 summarizes the quantitative accuracy of diffusion-based deep learning models and MRI-cGAN. The proposed CG-DDPM can achieve optimal results regarding MAE and PSNR for most of the translation tasks. Although IDDPM shows better PSNR values than CG-DDPM for T1 to T2, and FLAIR to T1 MRI synthesis, the statistical $p$-value tests indicate no significant difference in PSNR results. CG-DDPM achieves the optimal MSSIM for all types of MRI image generation. Figure 2 confirms the results that CG-DDPM can generate the most similar images as the ground truth regarding difference and resolution comparisons via MSSIM.

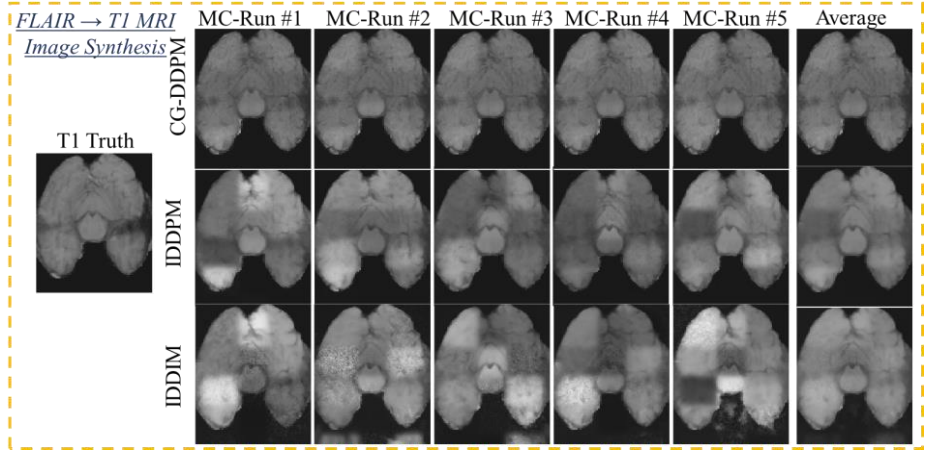

**Figure 3.** Ground truth MRIs, synthetic MRIs from five different runs and their average of the MC-based sampling of the DDPM models are presented column-wise. We present the proposed CG-DDPM in the first row, IDDPM in the second row, and IDDIM in the third row. More examples are shown in the Appendix. C.

### 3.4 Qualitative and quantitative results for consistency of CG-DDPM

We assessed the consistency of the synthetic result of the DDPM-based methods. Figure 3 shows that the synthetic results from five runs of the MC-based sampling and their average among the five runs. The proposed CG-DDPM shows higher consistency compared to the IDDPM and IDDIM among different runs. We further calculated the mean standard deviation of image intensities from the five runs as the uncertainty of DDPM-based deep learning models. In Figure 5, the CG-DDPM achieves the minimum uncertainties in all tasks, indicating superior consistency compared to other methods.

In addition, we ran the DDPM-based models for $n$ times to sample synthetic MRIs and denoted them as MC-$n$, where $n$ ranges from one to five. We calculated the MSSIM of the average results of all MC-$n$. Then, we normalized MSSIM (N-MSSIM) using the maximum SSIM value across all MC-$n$. This allowed us to observe how the quality of synthetic images changed with different numbers of runs in MC-based sampling. A slower convergence could indicate that more MC runs are needed to obtain the converged synthetic MRIs. In Figure 4, the N-MSSIM of CG-DDPM is converged with run times of one or two, which is faster than IDDPM and IDDIM. The standard deviations of N-MSSIM from different MC-$n$ results define the inconsistency. Figure 5 shows the inconsistency of diffusion-based models. CG-DDPM achieves the minimum inconsistency, indicating that the optimal and stable synthetic MRI can be obtained with minimum sampling runs.

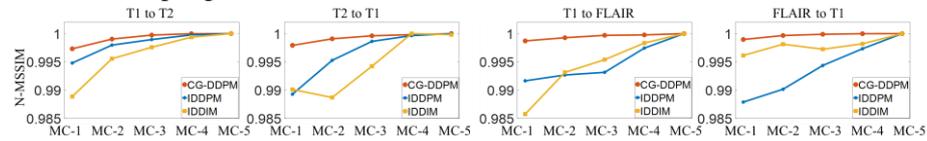

**Figure 4.** Sampling stability of the DDPM-based methods from four MRI translation tasks. MC-$n$ indicates the MC-based sampling averaged result using $n$ runs. Notice that the N-MSSIM does not represent the absolute quantitative performance (e.g., CG-DDPM's "1" is much higher than IDDIM's "1"), but a trend of the performance under different numbers of runs in the MC-based sampling.

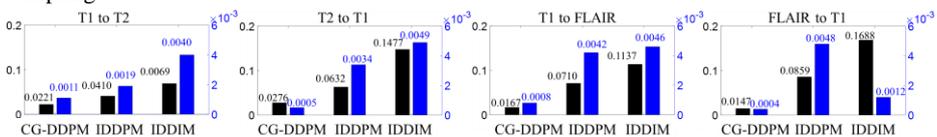

**Figure 5.** The quantitative robustness of the DDPM-based methods in four MRI translation tasks is presented through black and blue bars that represent sampling uncertainty and inconsistency, respectively, with corresponding values shown above the bars.

### 3.5 Discussion

The CG-DDPM network performed the best in terms of image quality, achieving the lowest MAE and highest MSSIM, as well as the second-best PSNR among all competing networks in all translation tasks. Compared to the IDDIM and MRI-cGAN networks, the CG-DDPM demonstrated statistically significant improvements (with p-values < 0.05) in reducing cumulative errors (lower MAE), improving accuracy of anatomical intensity, structure, and tissue contrast across multiple resolutions (higher

MSSIM), and reducing peak errors of synthetic MRI (smaller PSNR). Even compared to the IDDPM network, while the CG-DDPM had slightly worse PSNR, it showed better MAE and significantly better MSSIM, indicating its superior performance in synthesizing anatomical intensity, structure, and contrasts with smaller cumulative errors.

In terms of consistency, the CG-DDPM converges faster (Fig. 4) than the IDDPM and IDDIM to its optimal synthesis in the MC-based sampling. In addition, the CG-DDPM showed the lowest sampling uncertainty and inconsistency in all translation tasks. The CG-DDPM can generate accurate and stable synthetic MRIs with fewer sampling runs compared to other DDPM-based methods.

## 4    Conclusion

This work introduces a cycle-guided denoising diffusion probability model (CG-DDPM), which is the first 3D DDPM-based method for generating high-quality MRIs of a target modality using MRIs from a different modality. It uses two DDPMs to generate MRIs for each modality, and the random reverse latent noise from one DDPM is used to guide the other, which helps to generate matching MRIs in both modalities. The CG-DDPM can generate accurate MRIs from the target modality, and achieve state-of-the-art quantitative performance compared to the other networks. The CG-DDPM also reduce uncertainties in the synthetic MRIs compared to the other DDPM-based methods. As a result, the CG-DDPM can serve as a reliable tool for generating high-quality MRIs of a target modality given by MRIs from another modality, to facilitate MRI medical applications.

## Acknowledgments

This research is supported in part by the National Institutes of Health under Award Number R01CA215718, R56EB033332, and R01EB032680.